# The First Light Curve Solutions and Period Changes of EW Psc and HN Psc

F. Davoudi[1], A. Poro[1], F. Alicavus[3], J. Rahimi[2], E. Lashgari[1], M. Ghanbarzadehchaleshtori[1], A. Boudesh[1], S. Hoshyar[2], D. Asgarian[1], N. Pilghush[1], S. Modarres[2], S. Ebadi[2], A. Sojoudizadeh[1], S. Z. Azarniur[2], H. Karami[2], A. Dehghani Ghanatghestani.[1]

[1]The International Occultation Timing Association Middle East Section, Iran, info@iota-me.com
[2]The Seventh IOTA/ME Summer School of Astronomy, Astronomical Research Center, Qom, Iran
[3]Astrophysics Research Centre and Observatory, Çanakkale Onsekiz Mart University, Terzioğlu Kampüsü, TR-17020, Çanakkale, Turkey

**Abstract**
The first light curve solutions of the stars EW and HN in Pisces constellation are presented. Photometry, and its' periodic changes are calculated and discussed. The analysis of O-C diagram done by MCMC approach in OCFit code and the new ephemeris provided for two binary systems. The light curve solutions obtained. The results show that EW Psc is a near contact eclipsing binary system with a photometric mass ratio q = 0.587, and the fillout factor -0.034 and -0.018 for primary and secondary components, respectively. The solution results also show that the system HN Psc is a weak-contact W UMa eclipsing binary with a photometric mass ratio q = 0.853, and with fillout factor 5.6%. The light curves solutions required the cold spot accounting for the O'Connell effect.

**Introduction**
We present the first investigation of the photometry analysis of two W UMa binary systems, EW and HN that both are located in the Pisces constellation. The study includes light curve analysis and determination of new ephemeris for both systems and represents detailed modeling for the light curve fitting which leads to derive physical parameters of each system, also O-C analysis is accomplished for the aforementioned systems. Selected systems in terms of magnitude and period are very suitable for investigation. EW Psc (GSC 594-324) with the magnitude from 10.26 to 10.39, a period of 0.2412 day (Hoňková, K. et al., 2013); and HN Psc (GSC 2296-441) has magnitude from 10.75-10.95, with a period of 0.329539 day (Nelson, 2016).

**Observations**
The stars were observed for three nights on October 2019. The observation held at a personal observatory, Karaj, Iran (Long. 50.9380°E, Lat. 35.9545°N, Alt. 1623m) with a 16 inch Schmidt-Cassegrain telescope and a SBIG STF-8300M CCD. On these observation nights, we used a V filter with an exposure time of 40 seconds for each star. We obtained 202 images for EW Psc and 220 images for HN Psc in these observations. All the obtained CCD images were reduced by applying the bias, dark and flat field frames. Characteristics of target stars, comparisons and references are presented in Table 1, and Figures 1 and 2 show the observed and synthetic light curves of EW Psc and HN Psc.

Table 1. Specifications of the variables, comparisons, and references stars from SIMBAD[1].

| Type of star | Star name | Magnitude (V.) | RA. (2000) | Dec. (2000) |
|---|---|---|---|---|
| Variable | EW Psc | 10.37 | 00 01 11.48 | +09 04 41.48 |
| Comparison | GSC 594-1190 | 11.08 | 00 00 44.78 | +09 04 00.17 |
| Reference | GSC 594-318 | 11.39 | 00 00 41.80 | +09 09 48.16 |
| Variable | HN Psc | 10.55 | 01 29 47.92 | +33 03 35.66 |
| Comparison | GSC 2296-326 | 11.54 | 01 29 59.37 | +33 06 11.16 |
| Reference | GSC 2297-807 | 11.13 | 01 30 29.07 | +33 07 43.37 |

---

[1]http://simbad.u-strasbg.fr/simbad/

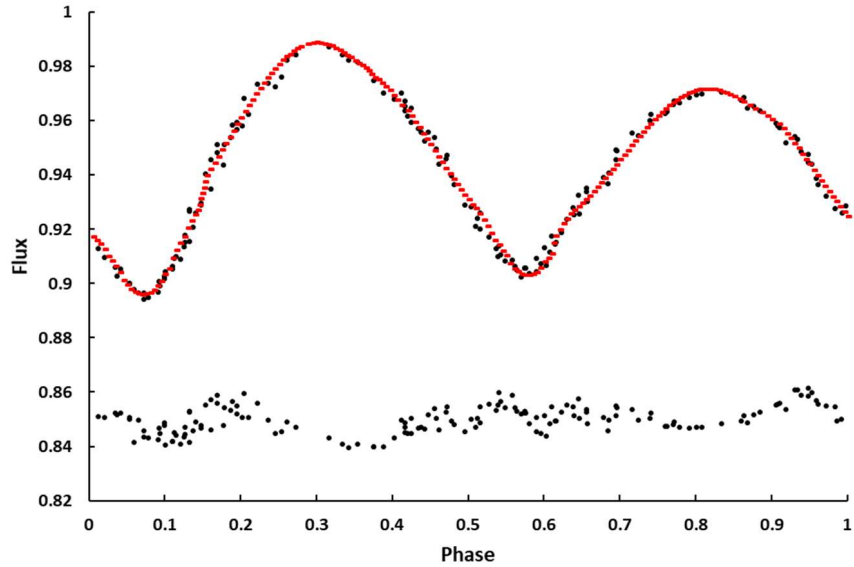

**Figure 1.** The observed light curves of EW Psc and synthetic light curves in V filter and residuals are plotted; with respect to orbital phase, shifted arbitrarily in the relative flux.

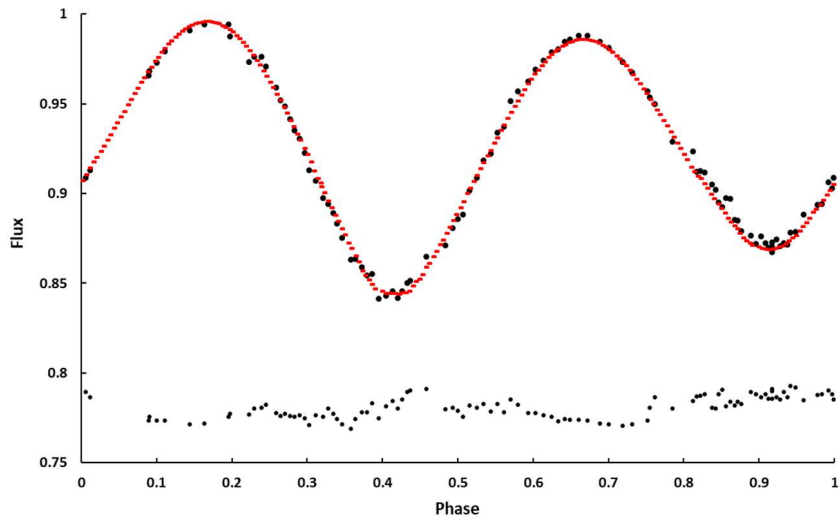

**Figure 2.** The observed light curves of HN Psc and synthetic light curves in V filter and residuals are plotted; with respect to orbital phase, shifted arbitrarily in the relative flux.

**Orbital period variations**

We collected all mid-eclipse times from the literature and obtained the individual mid-eclipse times for our observations of EW Psc and HN Psc; they were calculated using the Kwee and van Woerden (1956) method. We mentioned the reference ephemeris in $BJD_{(TDB)}$ for two systems in Table 2.

**Table 2.** The reference ephemeris of EW Psc and HN Psc.

| Star name | Reference Ephemeris | Reference |
|---|---|---|
| EW Psc | Min. I = $BJD_{(TDB)}$ 2456541.4878 + 0.241200 × E | Hoňková, K. et al., 2013 |
| HN Psc | Min. I = $BJD_{(TDB)}$ 2457365.5990 + 0.329539 × E | Nelson, 2016 |

We listed times of minima for EW Psc and HN Psc in Tables 3 and 4. Including mid-eclipse times in $BJD_{(TDB)}$ and their uncertainties, we took the averaged error amount for data points which authors didn't compute their uncertainties, minima types (p: primary or s: secondary), filters, epochs of these minima times, O-C values and the references of mid-eclipse times.

**Table 3. Times of light minimum of EW Psc.**

| Min (BJD$_{TDB}$) | Error | Min Type | Epoch | O-C | Reference |
|---|---|---|---|---|---|
| 2453958.9381 | | p | -10707 | -0.0209 | IBVS 5806 |
| 2453975.9434 | | s | -10636.5 | -0.0202 | IBVS 5806 |
| 2455815.4757 | 0.0010 | p | -3010 | -0.00002 | OEJV 0160 |
| 2456541.4878 | 0.0006 | p | 0.0 | 0.0 | OEJV 0160 |
| 2457244.4674 | | s | 2914.5 | 0.0020 | OEJV 0179 |
| 2457257.4978 | | s | 2968.5 | 0.0076 | OEJV 0179 |
| 2457284.3872 | | p | 3080 | 0.0032 | IBVS 6196 |
| 2457616.5252 | | p | 4457 | 0.0087 | IBVS 6244 |
| 2457968.5651 | | s | 5916.5 | 0.0172 | B.R.N.O. |
| 2458041.4080 | | s | 6218.5 | 0.0177 | B.R.N.O. |
| 2458041.5228 | | p | 6219 | 0.0119 | B.R.N.O. |
| 2458042.3733 | 0.0004 | p | 6222.5 | 0.0182 | B.R.N.O. |
| 2458042.4895 | 0.0007 | s | 6223 | 0.0138 | B.R.N.O. |
| 2458043.4526 | | p | 6227 | 0.0121 | B.R.N.O. |
| 2459142.3659 | 0.0001 | p | 10783 | 0.0180 | This study |
| 2459142.4865 | 0.0002 | s | 10783.5 | 0.0180 | This study |

**Table 4. Times of light minimum of HN Psc.**

| Min (BJD$_{TDB}$) | Error | Min Type | Epoch | O-C | Reference |
|---|---|---|---|---|---|
| 2456584.3322 | 0.0032 | p | -2371 | 0.0702 | OEJV 0168 |
| 2456584.3322 | 0.0004 | s | -2224.5 | 0.0702 | IBVS 6092 |
| 2456632.6212 | | p | -1072 | 0.0817 | OEJV 0179 |
| 2457012.3883 | | p | -1072 | 0.0550 | OEJV 0179 |
| 2457012.4128 | | p | -324 | 0.0795 | IBVS 6164 |
| 2457365.5990 | | p | 0.0 | 0.0 | IBVS 6164 |
| 2457258.8702 | | s | 1103.5 | 0.0418 | OEJV 0194 |
| 2457729.2586 | | s | 1908.5 | 0.0133 | OEJV 0194 |
| 2457994.5210 | | p | 1984 | -0.0031 | IBVS 6244 |
| 2458019.3982 | | p | 1996 | -0.0061 | IBVS 6244 |
| 2458023.3539 | | s | 1996.5 | -0.0049 | IBVS 6244 |
| 2458023.5129 | 0.0003 | p | 2221 | -0.0106 | B.R.N.O. |
| 2458097.3298 | 0.0002 | s | 2220.5 | -0.0105 | B.R.N.O. |
| 2458403.2863 | 0.0002 | p | 3149 | -0.0309 | This study |
| 2458403.4521 | 0.0002 | s | 3149.5 | -0.0299 | This study |

The corresponding O-C diagrams were plotted in Figure 3 for EW Psc and in Figure 4 for HN Psc. Primary minimums were displayed by full circles and the secondary minimums by empty circles.

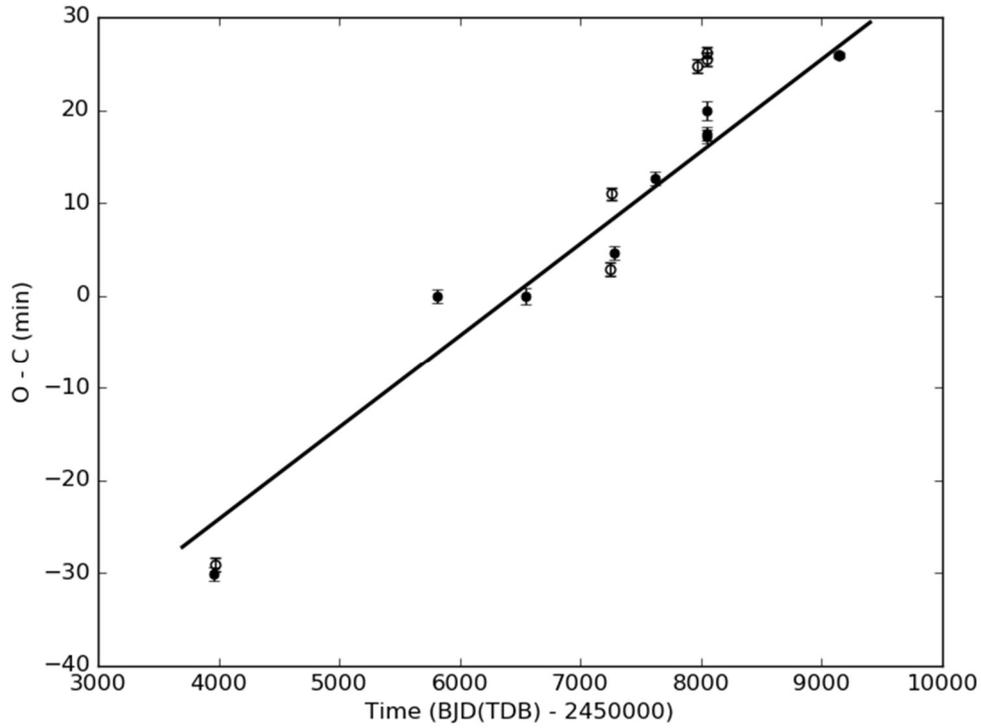

Figure 3. The O–C diagram of EW Psc with the derived line fitted on the data. The primary mid-eclipse times were displayed by full circle and the secondary mid-eclipse times by empty circle.

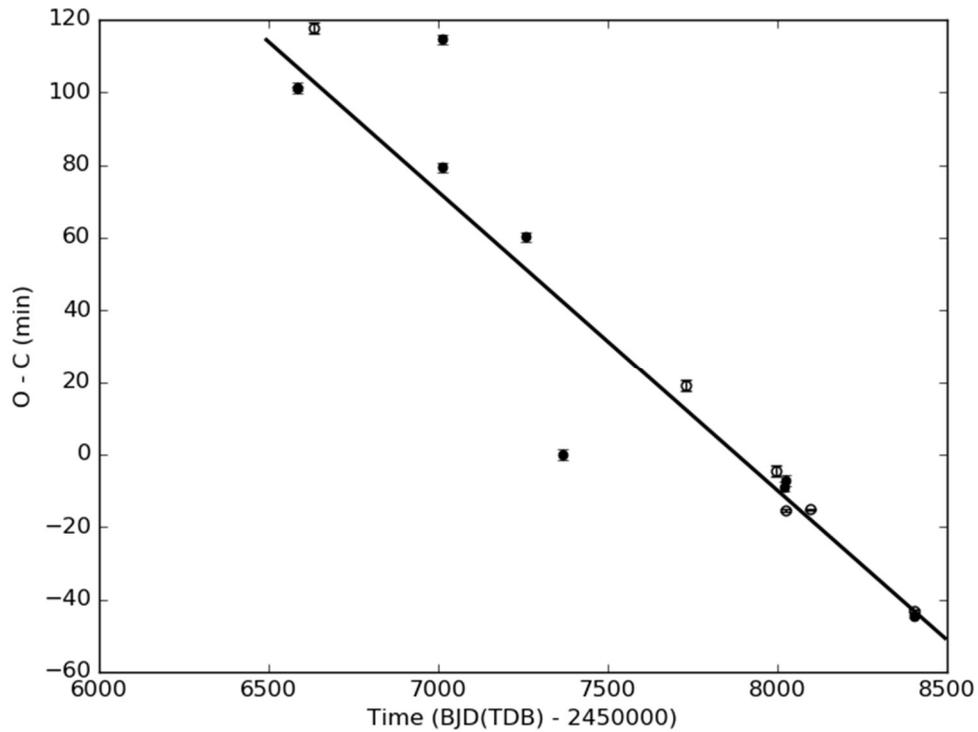

Figure 4. The O–C diagram of HN Psc with the derived line fitted on the data. The primary mid-eclipse times were displayed by full circle and the secondary mid-eclipse times by empty circle.

We fitted all mid-eclipse timings with a line, using the Monte Carlo Markov Chain (MCMC) approach in OCFit code (Gajdoš & Parimucha., 2019). Employing 1000000 iterations for the O-C diagram in our MCMC runs (1000000 MC steps); Number of first MC steps which will be rejected from the calculation and the size of one block for binning regarded as 5000 and 10 respectively. Confidence interval graphs for fitted parameters and histograms of parameters determined by the MCMC simulation depicted in Figures 5 and 6 for two binary systems. In confidence interval graphs, the red point is solution and regions display $1\sigma$ and $2\sigma$.

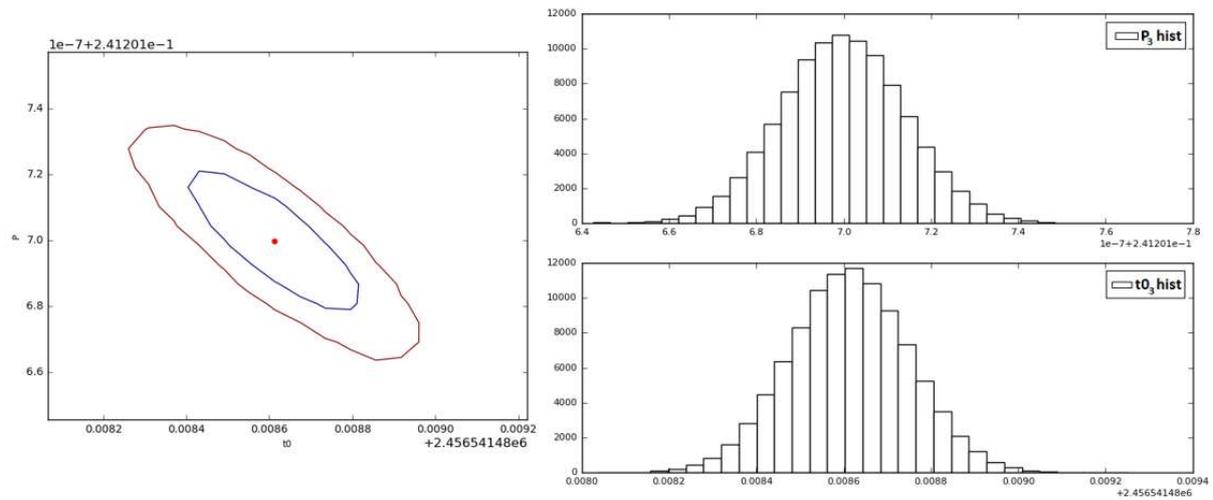

Figure 5. Confidence interval graph and histograms of fitted parameters determined by the MCMC simulation for EW Psc.

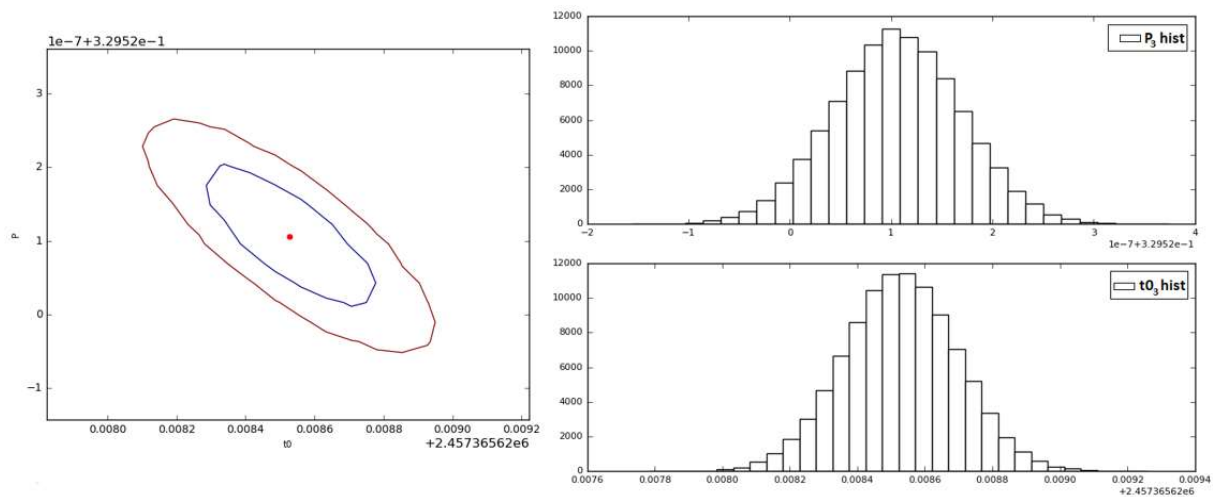

Figure 6. Confidence interval graph and histograms of fitted parameters determined by the MCMC simulation for HN Psc.

We determined new ephemeris of systems for the primary minimum in Table 5; E is the number of cycles after the reference epoch, and the values in brackets show the errors of new mid-eclipse time and new period.

Table 5. The New ephemeris of EW Psc and HN Psc.

| Star name | New Ephemeris |
| --- | --- |
| EW Psc | $T_0$ = BJD$_{(TDB)}$ 2456541.4886 (±0.0001) + 0.241201 (±0.0000001) × E |
| HN Psc | $T_0$ = BJD$_{(TDB)}$ 2457365.6285 (±0.0001) + 0.329520 (±0.0000006) × E |

We computed the variation of period of EW Psc $\frac{dp}{dt} = 2.371318 \times 10^{-3}\ days/year$ and realized it was increased from 2013 to 2020. As well there is a decrease in the period of $\frac{dp}{dt} = -6.402479 \times 10^{-4}\ days/year$ for HN Psc from 2015 to 2020. However, more observations are needed to analyze period changes in these systems.

**Light Curve Analyze**

The light curves of EW and HN Psc were analyzed based on the binary star model of Wilson & Devinney (1971, hereafter WD) and physical parameters were obtained by Phoebe 0.32 software. The following parameters were fixed in the analysis of both binary systems: the temperature of both binary systems which was obtained from Gaia DR2[2] data, the gravity darkening exponents of $g_1 = g_2 = 0.32$ (Lucy, 1967), the bolometric albedo coefficients of $A_1 = A_2 = 0.50$ (Rucinski, 1969), and linear limb darkening coefficients taken from tables published by Van Hamme (1993).

Based on the light curves, and the placement of the primary and secondary minimums, we could detect that the primary star in EW Psc is hotter than the secondary, and for HN Psc the secondary star is hotter than primary. After the initial analysis, we chose mod 2 for the EW Psc and Mod 3 for the HN Psc.

The q search method was used to obtain the mass ratio. So we find the minimum sum of squared residuals of W-D fit, $\sum(O-C)^2$. As a result, the range of fixed mass ratio for EW Psc from 0.1 to 1.3, and for HN Psc from 0.1 to 1.5 is shown in Figure 7. The q search diagram for EW Psc shows two main minima and we examine both in our light curve solutions. Therefore, a minimum value of $\sum(O-C)^2$ was initially achieved at q = 0.587 for EW Psc and q = 0.853 for HN Psc.

Based on these initial analyzes and using Phoebe 0.32 software, the parameters obtained from the light curve analysis of EW and HN Psc are presented in Tables 6.

The fillout factors ($f$) calculated from the output parameters of the light curve solutions. This is a measure of the amount of contact between components in the binary system. In the Formula (1), the Ω is the surface of the common envelope, Ω(L$_1$) is the potential at the inner Lagrangian surface, and Ω(L$_2$) is the outer Lagrangian surface.

$$f = \frac{\Omega(L_1) - \Omega}{\Omega(L_1) - \Omega(L_2)} \qquad (1)$$

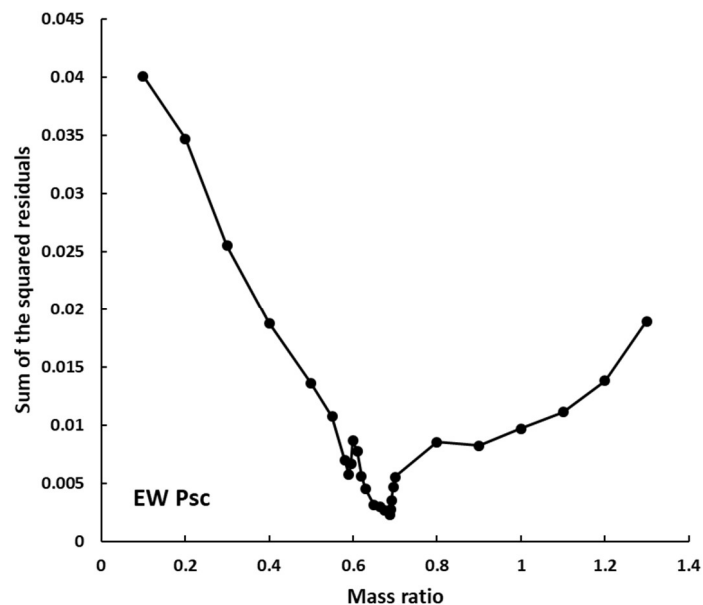

---

[2] https://www.cosmos.esa.int/web/gaia/dr2

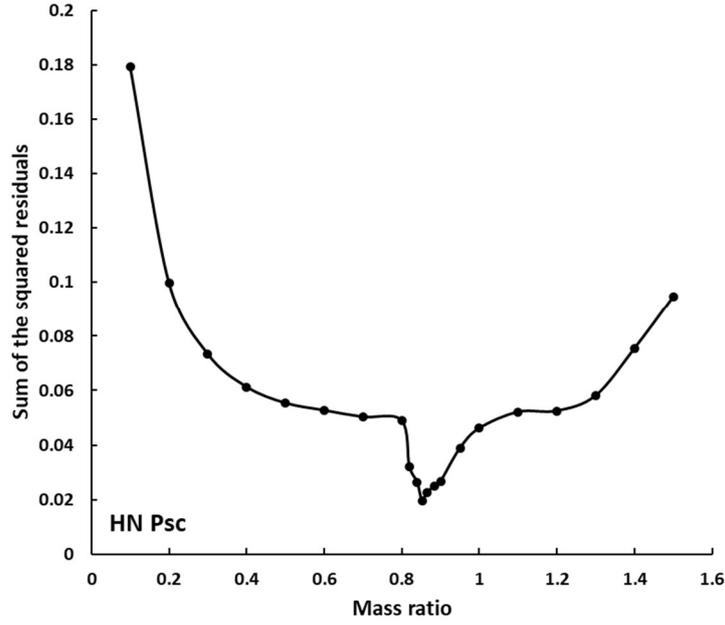

**Figure 7.** Sum of the squared residuals as a function of the mass ratio; at the top is EW Psc and at the bottom is HN Psc.

Table 6. Photometric solutions of EW Psc and HN Psc.

| EW Psc | | HN Psc | |
|---|---|---|---|
| Parameter | Value | Parameter | Value |
| $T_1$ (K) | 5355 | $T_1$ (K) | 5124(9) |
| $T_2$ (K) | 5128(202) | $T_2$ (K) | 5743(177) |
| $\Omega_1$ | 3.152(257) | $\Omega_1$ | 3.48(18) |
| $\Omega_2$ | 3.099(249) | $\Omega_2$ | 3.48(18) |
| i (deg) | 51.18(1.2) | i (deg) | 51.12(87) |
| $q=m_2/m_1$ | 0.587(17) | $q=m_2/m_1$ | 0.853(13) |
| $l_1/(l_1+l_2)$ | 0.743(34) | $l_1/(l_1+l_2)$ | 0.372(15) |
| $l_2/(l_1+l_2)$ | 0.256(13) | $l_2/(l_1+l_2)$ | 0.627 |
| $A_1=A_2$ | 0.5 | $A_1=A_2$ | 0.5 |
| $g_1=g_2$ | 0.32 | $g_1=g_2$ | 0.32 |
| $f_1$ | -0.034 | $f_1$ | 0.056 |
| $f_2$ | -0.018 | $f_2$ | 0.056 |
| $r_{1(mean)}$ | 0.426 | $r_{1(mean)}$ | 0.397 |
| $r_{2(mean)}$ | 0.342 | $r_{2(mean)}$ | 0.369 |
| Phase shift | 0.075(1) | Phase shift | -0.080(2) |
| Primary Spot: | | Primary Spot: | |
| Colatitude$_{sopt}$ (deg) | 34 | Colatitude$_{sopt}$ (deg) | 165 |
| Longitude$_{spot}$ (deg) | 66 | Longitude$_{spot}$ (deg) | 68 |
| Radius$_{spot}$ (deg) | 22 | Radius$_{spot}$ (deg) | 67 |
| $T_{spot}/T_{star}$ | 0.91(2) | $T_{spot}/T_{star}$ | 0.93(1) |
| Secondary Spot: | | | |
| Colatitude$_{sopt}$ (deg) | 87 | | |
| Longitude$_{spot}$ (deg) | 66 | | |
| Radius$_{spot}$ (deg) | 19 | | |
| $T_{spot}/T_{star}$ | 0.90(3) | | |

The mean fractional radii of components were calculated with the formula, $r_{mean} = (r_{pole} \times r_{side} \times r_{back})^{1/3}$. The position of the stars in the binary systems are shown in Figure 8.

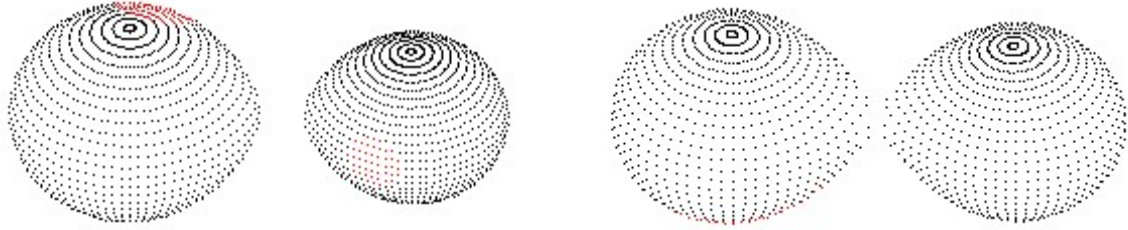

**Figure 8.** The position of the stars. On the left is EW Psc in phase 0.320, and on the right HN Psc in phase 0.165.

The O'Connell effect can be recognized clearly in the light curves of our observations (O'Connell, 1951). Which caused us to place two cold spots on the primary and secondary components of the EW Psc, and one cold spot on the primary component of HN Psc. This effect is more visible in the EW Psc. Table 7 presents EW and HN Psc's characteristic parameters of the light curves, along with the maximum difference.

**Table 7. Characteristic parameters of the light curves in V filter.**

| Part of LC. | EW Psc | HN Psc |
|---|---|---|
| MaxI- MaxII | -0.020 | 0.008 |
| MaxI - MinII | -0.100 | -0.176 |
| MaxI - MinI | -0.110 | -0.141 |
| MinI - MinII | 0.010 | -0.035 |

**Conclusion**

In this study we presented a new observation and light curves for EW Psc and HN Psc and calculated ephemeris for determining new the time of minima. The first photometric solutions of these binary systems were studied. The light curve solutions were done by Phoebe 0.32 software based on the Wilson & Devinney code. The analysis of O-C diagram done by OCFit code. We fitted all mid-eclipse timings with a line, using the MCMC approach. Confidence interval graphs for fitted parameters and histograms of parameters determined by the MCMC simulation displayed.

The light curve analysis of EW Psc has shown the system to be a near-contact eclipsing binary with the negative fillout factor -0.034 and -0.018 for primary and secondary components, respectively. The variation of period of EW Psc $\frac{dp}{dt} = 2.371318 \times 10^{-3}\ days/year$ shows that it has been increasing from 2013 to 2020. The difference in their mean effective temperatures is only of the order of 200$^K$. The light curve solutions for HN Psc shows that this is a weak-contact binary system with a fillout factor 5.6%. The O-C diagram of HN Psc indicates a decrease in the period of $\frac{dp}{dt} = -6.402479 \times 10^{-4}\ days/year$ from 2015 to 2020. Based on the phase-flux light curve for HN Psc, the secondary component is hotter than primary and they have different effective temperatures in the order of 600$^k$.

**Acknowledgments**

This manuscript was prepared by the International Occultation Timing Association Middle East section (IOTA/ME) and with participants at 7[th] International Astronomy Summer School, ARC, Qom, Iran between 13 to 16 August, 2019.

**References**

[1] B.R.N.O., Project-Variable Star and Exoplanet Section of Czech Astronomical Society, Contributions #41. <http://var2.astro.cz/EN/brno/index.php>.
[2] Gajdoš, P. and Parimucha, S., 2019. New tool with GUI for fitting OC diagrams. Open European Journal on Variable Stars, 197, p.71.
[3] Hoňková, K., Juryšek, J., Lehký, M., Šmelcer, L., Trnka, J., Colazo, C., Guzzo, P., Mina, F.D., Quinones, C., Taormina, M. and Melia, R., 2013. BRNO Contributions# 38 Times of minima. Open European Journal on Variable Stars, 0160.
[4] Hoňková K., et al., 2015. B.R.N.O. Contributions #39 Times of minima. Open European Journal on Variable Stars, 0168.

[5] Hübscher J., 2017. BAV-Results of observations - Photoelectric Minima of Selected Eclipsing Binaries and Maxima of Pulsating Stars. Information Bulletin on Variable Stars, 6196, #1.

[6] Juryšek J., et al., 2017. B.R.N.O. Contributions #40 Times of minima. Open European Journal on Variable Stars, 0179.

[7] Krajci T., 2007. Photoelectric Minima of Some Eclipsing Binary Stars. Information Bulletin on Variable Stars, 5806, #1.

[8] Kwee, K. and Van Woerden, H., 1956. A method for computing accurately the epoch of minimum of an eclipsing variable. Bulletin of the Astronomical Institutes of the Netherlands, 12, p.327.

[9] Lucy, L.B., 1967. Gravity-darkening for stars with convective envelopes. Zeitschrift fur Astrophysik, 65, p.89.

[10] Nelson, R. H., 2016. Timings of minima of eclipsing binaries. Information Bulletin on Variable Stars, 6164, #1.

[11] Nelson R. H., 2014. CCD Minima for Selected Eclipsing Binaries in 2013. Information Bulletin on Variable Stars, 6092, #1.

[12] O'Connell, D.J.K., 1951. The so-called periastron effect in eclipsing binaries. Monthly Notices of the Royal Astronomical Society, 111(6), pp.642-642.

[13] Pagel L., 2018. BAV-Results of observations - Photoelectric Minima of Selected Eclipsing Binaries and Maxima of Pulsating Stars. Information Bulletin on Variable Stars, 6244, #1.

[14] Rucinski, S.M., 1969. The proximity effects in close binary systems. II. The bolometric reflection effect for stars with deep convective envelopes. Acta Astronomica, 19, p.245.

[15] Van Hamme, W., 1993. New limb-darkening coefficients for modeling binary star light curves. The Astronomical Journal, 106, pp.2096-2117.